\documentstyle{article}
\begin{document}
\title{ On the applicability of Sato's equation to Capacitative RF  
Sheaths  }
\author{J.Balakrishnan \thanks{E-mail : janaki@serc.iisc.ernet.in, ~janaki@hve.iisc.ernet.in } ~ 
\& G.R.Nagabhushana \thanks{E-mail : grn@hve.iisc.ernet.in } \\
Department of High Voltage Engineering, Indian Institute of Science, 
\\ 
Bangalore -- 560 012, India.  }
\date{\mbox{\ }}
\maketitle
\vspace{1.5cm}
\begin{flushright}
P.A.C.S. numbers ~~: ~52.50.Dg, ~52.90.+z, ~52.40.Hf, ~77.22.Jp~~~
\end{flushright}
\vspace{1.5cm} 
\newpage
\begin{abstract} 

We show that the time dependent version of Sato's equation, when 
applied to capacitative rf sheaths is no longer independent of the 
electric field of the space charge, and discuss the use of the equation for a specific sheath model. 
\end{abstract}
\newpage

\subsection*{1. ~Introduction}
A few years ago, Sato derived an expression for the current flowing in 
the external circuit due to the motion of charged particles in the gap 
within a discharge tube [1], for a constant voltage applied to the 
gap. Very recently, this work was generalised by Morrow \& Sato [2] to 
include time-dependent applied voltages. They used an energy balance 
equation to obtain their results, which in both cases were independent 
of the space charge effects. \\
While their methods and inferences drawn thereby are striking in their 
elegance and simplicity, one must exercise caution while applying 
them, for in certain situations, these can lead to results which are 
quite the reverse of the inferences drawn in their papers, as we show 
here. \\ 
In a high voltage radio frequency discharge tube, there is a high 
concentration of positive ions at the electron-depleted dark space 
region adjoining the electrode, and a consequent build-up of field 
distortion, so that one expects a time and frequency dependence of the 
conduction process. The build-up of field distortion can therefore be 
modelled by a two-layer capacitor of the Maxwell-Wagner type [3]. \\
The plasma state is characterised by equal numbers of positive and 
negative charges, but because of diffusion effects and recombinations 
on boundary surfaces in the discharge tube, there is charge depletion 
in the adjoining gas phase, resulting in the formation of a thin 
sheath. Electrons diffuse fastest, since they are lightest and have 
high energies, and they leave behind them a surplus of positive 
charge and a plasma potential which is positive relative to the walls. 
Since there is a larger number of charged particles in the central 
plasma regions of the tube, and hence better conductivity there, 
almost all of the potential drop occurs across the sheath.  
The clear difference in the magnitudes of the potential at the plasma 
and at the sheath, leads one to describe the plasma-sheath system in a 
discharge tube by a two-layered capacitor, with a dielectric 
coefficient $\epsilon_{sh}$ ~ for the sheath region. \\

It is shown in this work that such a representation of the 
plasma-sheath system leads to terms in Sato's equation, which depend 
upon the sheath potential and field. We calculate the non-zero contributions to the current in the external circuit from the space charge field and the sheath capacitance for a symmetric RF discharge for a simplified electrode geometry, in the case of the Godyak-Lieberman (GL) theory [4,5] for RF sheaths, using this energy balance method.

\subsection*{2. ~The energy balance method of Sato \& Morrow}
The continuity equations which describe the development in time $t$ of 
the space charge in a gap in air are [6]:
\begin{eqnarray}
\frac{\partial n_e}{\partial t} &=& n_e\alpha{\bf v}_e -n_e\eta{\bf 
v}_e - n_en_p\beta_1 - \nabla\cdot( n_e{\bf v}_e - D_e\nabla n_e ) 
\nonumber\\
\frac{\partial n_p}{\partial t} &=& n_e\alpha{\bf v}_e - n_en_p\beta_1 
- n_nn_p\beta_2- \nabla\cdot( n_p{\bf v}_p - D_p\nabla n_p ) 
\nonumber\\
\frac{\partial n_n}{\partial t} &=& n_e\eta{\bf v}_e - n_nn_p\beta_2 - 
\nabla\cdot( n_n{\bf v}_n - D_n\nabla n_n ) 
\end{eqnarray}
where $n_e, n_p$ and $n_n$ denote the electron, positive ion and 
negative ion densities respectively, ${\bf v}_e, {\bf v}_p$ and ${\bf 
v}_n$ denote the drift velocity vectors of the electrons, positive 
ions \& negative ions respectively, $\alpha$ is the electron 
ionization coefficient, $\eta$ the attachment coefficient, and 
$\beta_1$ and $\beta_2$ are the electron-positive ion and negative 
ion-positive ion recombination coefficients respectively. $D_e, D_p$ 
and $D_n$ are the diffusion coefficients for the electron, positive 
ion and negative ion respectively.\\
Combining these, one obtains the equation for the net space charge 
density $\rho$ :
\begin{equation} 
\frac{\partial\rho}{\partial t} = e\frac{\partial(n_p -n_e 
-n_n)}{\partial t} = -e \nabla\cdot\Gamma  
\end{equation} 
where $e$ is the electron charge, and the total particle flux $\Gamma$ 
is given by 
\begin{equation} 
\Gamma = n_p{\bf v}_p - n_n{\bf v}_n - n_e{\bf v}_e - D_p\nabla n_p + 
D_n\nabla n_n + D_e\nabla n_e 
\end{equation} 
One starts with these basic equations to study the electrodynamics of the charged particles in a discharge tube. 
In an RF discharge tube, since most of the potential drop in the gap occurs across the sheath, 
we treat the plasma-sheath system as a two-layer capacitor, with a 
dielectric constant $\epsilon_{sh}$ for the sheath and the dielectric 
constant of free space $\epsilon_0$ for the plasma region. \\ 
The total current density $J$ in the gap is 
\begin{equation}
J = e\Gamma + \frac{\partial D}{\partial t}
\end{equation} 
where the electric displacement $D$ relates to the local field $E$ 
through the effective complex dielectric constant $\epsilon$ of the 
gap: 
\begin{equation}
D = \epsilon E 
\end{equation} 
This can be separated into a part $D_{sh}$ describing the sheath, and 
a plasma part $D_p$:
\begin{equation}
D = \epsilon_0E_p + \epsilon_{sh}E_{sh} = D_p + D_{sh} 
\end{equation}
$D$ obeys Poisson's equation : 
\begin{equation}
\nabla\cdot D = \rho = \epsilon_0\nabla\cdot E_p + \nabla\cdot 
(\epsilon_{sh}E_{sh})
\end{equation}
while $E_p$ satisfies Laplace's equation:
\begin{equation}
\nabla\cdot E_p = 0 
\end{equation} 
The plasma and sheath electric fields $E_p$ and $E_{sh}$ are related 
to their respective potential distributions through 
\begin{equation}
E_p = -\nabla\psi_p  ~~~ {\rm and} ~~~  E_{sh} = -\nabla\psi_{sh}  .
\end{equation}
The energy balance equation can be used to relate the applied 
potential $V_a$ and the current $I$ in the external circuit, to the 
current density and the local field in the gap through the volume 
integral $\int_V dv$ over the discharge space:
\begin{equation}
V_aI = \int_V J.E dv 
\end{equation}
Separating out the plasma and the sheath electric fields as: 
\begin{equation}
E = E_p + E_{sh} 
\end{equation}
and making use of (4), one obtains:\\
\begin{eqnarray}
V_aI &=& \int_V [ e\Gamma + \frac{\partial D}{\partial t} ]\cdot E_p 
dv  + \int_V [ e\Gamma + \frac{\partial D}{\partial t} ]\cdot E_{sh} 
dv \nonumber\\ 
&=& \int_V [ e\Gamma + \frac{\partial D}{\partial t} ]\cdot E_p dv - 
\int_V \psi_{sh}(x,t)\frac{\partial\rho(x,t)}{\partial t} dv - \int_S 
e\Gamma\psi_{sh}(x,t)\cdot dS \nonumber\\
& &~~~~~~~~~~~~~~~~~~~~~~~~~~~~~~~~~~~~~~~~~~~~+ \int_V \frac{\partial D}{\partial 
t}\cdot E_{sh} dv
\end{eqnarray}
where we have made use of the second of equations (9), performed an 
integration by parts, $\int_S dS$ representing a surface integration 
over the closed surface of the discharge space, and then used (2).\\   
The second term in the right hand side of (12) can be rewritten using 
(7), and after again performing an integration by parts, this term can 
be written as:
\begin{equation}
\int_V \psi_{sh}(x,t)\frac{\partial\rho(x,t)}{\partial t} dv = \int_V 
\frac{\partial D}{\partial t} E_{sh} dv + \int_S 
\psi_{sh}(x,t)\frac{\partial D(x,t)}{\partial t} dS
\end{equation}
Next, we substitute (13) and (6) back into (12) to obtain:
\begin{equation}
V_aI = \int_V [ e\Gamma + \epsilon_0\frac{\partial E_p}{\partial t} 
]\cdot E_p dv + \int_V \frac{\partial D_{sh}}{\partial t} E_p dv - 
\int_S ( e\Gamma + \frac{\partial D}{\partial t} )\psi_{sh} dS
\end{equation}

The first two terms within the first volume integral on the right hand 
side of (14) constituted the final form of Sato's equation derived in 
[2] for a time-dependent applied voltage. The boundary condition 
chosen in [2] was, that $\psi_{sh}$ was set to zero on both 
electrodes, and $\psi_p$ was set to zero on one electrode and to the 
applied voltage $V_a$ on the other. The authors in [2] obtained the 
gap capacitance for a simple system from the second of the two terms 
in the first volume integral.

On the other hand, by ascribing a non-trivial dielectric constant to 
the sheath, we have obtained terms additional to those obtained in 
[2], and these extra terms {\em depend} upon the sheath field and 
potential.\\ 

\subsection*{3. ~Capacitative RF discharges}
Low pressure plasma chambers are widely used in the material 
processing industry, such as in the fabrication of semiconductor 
wafers, and in reactive ion etching. Many applications need high ion 
energies which are generated by biasing the substrate with a radio 
frequency (RF) current source [7,8]. Proper understanding of the 
electrodynamics involved in the sheath and near the plasma-sheath 
boundary is thus highly desirable.\\ 
We consider a low density, low pressure plasma where a single power 
supply generates both the discharge and the RF sheath. In this high  
frequency regime, the sheaths are primarily capacitative in nature. 
The sheath is assumed to be collisionless. As is well known [4,5,9,10], the analysis of sheath models depends on the ratio of the applied RF frequency $\omega$ to the ion plasma frequency $\omega_{pi}$ ~:~~ when  ~$\omega << \omega_{pi}$, ~ the ions cross the sheath quickly and can 
instantaneously adjust to the applied field, and the properties of the 
sheath at different times of the RF cycle are identical to those of a 
dc sheath having a potential given by its instantaneous value (RF plus 
dc). However, when ~$\omega >> \omega_{pi}$, ~ the inertia of the ions 
prevents them from adjusting to the applied field, and they cannot 
respond to its time variation. Then the ions cross the sheath in many 
RF periods and they respond only to the  dc field  -- their dynamics 
is governed by the time-averaged field in the sheath.\\    
In the Godyak-Lieberman (GL) theory [4,5] (which we consider now), 
valid in the low density, high frequency, high current regime [11], 
the ions are assumed to react only to the dc fields and not to the RF 
fields. Also it is assumed that the transit time for the ions across 
the sheath is large compared to the oscillation time. These 
assumptions lead the GL theory to predict monoenergetic ions impinging 
on the substrate.\\ 
The ions are assumed to enter the sheath with a Bohm presheath 
velocity [12] ~ $v_B = \frac{eT}{m_i}$ ~ where $e$ is the ion charge, $T_e$ 
denotes the electron temperature in volts and $m_i$ is the ion mass. 
The ion sheath-plasma boundary is taken to be stationary, and it is 
assumed that the electrons being inertialess, respond to the 
instantaneous field. \\
The GL theory holds in the regime in which the applied RF voltage is 
very large compared to $T_e$, so that one can assume that the electron 
Debye length $\lambda_D$ everywhere within the sheath is much smaller 
than the thickness of the ion sheath $s_m$, implying that the  
electron density drops sharply from $n_e = n_i$ at the boundary with 
the plasma to $n_e = 0$ in the sheath (at the electrode side). The 
electron sheath penetrates into the ion sheath for a distance $s(t)$ 
from the plasma-ion sheath boundary at $x=0$, and oscillates between a 
maximum thickness of $s_m$ and a minimum thickness which is a few 
Debye lengths distant from the electrode, so that the electron sheath 
thickness is effectively $s_m - s(t)$.\\
We follow here, the analysis given in [4,5], but modified to include a finite dielectric constant $\epsilon_{sh}$ for the sheath. \\
The ion flux is conserved at the plasma-ion sheath boundary. This is 
expressed by:
\begin{equation}
n_iv_i = n_0v_B 
\end{equation}
where we denote the ion density $n_i(x)$ at the plasma-ion sheath 
boundary by $n_0$, and $v_i$ is the ion velocity. From energy 
conservation, one obtains:
\begin{equation}
\frac{1}{2}m_iv_i^2 = \frac{1}{2}m_iv_B^2 - e\bar\psi_{sh}(x)
\end{equation}
where $\bar\psi_{sh}(x)$ is the time averaged potential within the 
sheath. Combining (15) and (16), one obtains for the ion density:\\
\begin{equation}
n_i(x) = n_0{\bigl(1-\frac{2}{T_e}\bar\psi_{sh}(x)\bigr)}^{-1/2} 
\end{equation}
In the GL theory, a spatially uniform, sinusoidal RF current density 
is assumed to pass through the sheath:
\begin{equation}
J_{RF}(t) = - J_0\sin\omega t 
\end{equation}
This current is carried by the electrons : ~$J = -ne\frac{ds}{dt}$,~ 
$n$ being the electron density in the bulk plasma at the sheath edge. 
As these electrons oscillate away from the electrode, they leave 
behind a positive space charge space, pulling the positive ions there.
At the electron sheath boundary, the displacement current given by 
(18) must be equated to the conduction current, for continuity :
\begin{equation}
-en_i(s)\frac{ds}{dt} = - J_0\sin\omega t  
\end{equation}
The instantaneous electric displacement $D_{sh}(x,t)$ within the 
sheath is then given by:
\begin{eqnarray}
\frac{\partial D_{sh}}{\partial x} &=& en_i(x) , ~~~~~~ x>s(t) 
\nonumber\\
&=& 0 , ~~~~~~~~~~~~ x<s(t)
\end{eqnarray}
where $D_{sh}(x,t)= \epsilon_{sh}(x,t)E_{sh}(x,t)$ .\\  
The time-averaged electric displacement and potential are given by:
\begin{eqnarray}
\frac{d\bar D_{sh}}{dx} &=& e(n_i(x) - \bar n_e(x)) \nonumber\\ 
\frac{d\bar\psi_{sh}}{dx} &=& -\bar E_{sh} 
\end{eqnarray}
$\bar n_e(x)$ being the time averaged electron density within the 
sheath.\\
The electric displacement field within the sheath can be found by 
integrating (20):
\begin{equation}
D_{sh} = e\int^x_{s(t)} n_i(\xi) d\xi 
\end{equation}
This is done with the help of eqn.(19). Integrating its left hand side 
between the limits $0$ and $s$, and the right hand side between the 
limits $0$ and $\omega t$, one gets:
\begin{equation}
e\int^s_0 n_i(\xi) d\xi = \frac{J_0}{\omega}(1- \cos\omega t) 
\end{equation}
Using (23) in (22), one obtains\\
\begin{eqnarray}
D_{sh}(x,\omega t) &=& e\int^x_{s(t)} n_i(\xi) d\xi = ~e\int^x_0 
n_i(\xi) d\xi - e\int^{s(t)}_0 n_i(\xi) d\xi ~, ~~~ x > s(t) 
\nonumber\\
&=& 0 ~, ~~~~~~~~~~~~~~~~~~~~~~~~~~~~~~~~~~~~ 
~~~~~~~~~~~~~~~~~~~~~~~~~~~x<s(t)
\end{eqnarray}
Since the GL theory is valid for the high frequency regime, one is 
interested in the time-averaged quantities. These can be found from 
$s(t)$.  Lieberman denotes by  $2\phi$,  the phase during which 
$x>s(t)$ : ~ then for $x\approx 0$, ~ $2\phi\approx 0$ , and for 
$x\approx s_m$, ~ $2\phi\approx 2\pi$. Because $n_e(x,t)=0$ during the 
part of the RF cycle when $x>s(t)$, he writes:
\begin{equation}
\bar n_e(x) = \bigl(1-\frac{2\phi}{2\pi}\bigr)n_i(x)
\end{equation}
so that ~$-\phi < \omega t < \phi$ , for ~$x>s(t)$,~ and ~$\omega t = 
\pm\phi$ ~for ~$x=s(t)$.\\
Then, from (23) and (24), one obtains:
\begin{eqnarray}
\bar D_{sh}(x) &=& \frac{1}{2\pi}\int_{-\phi}^{+\phi} D_{sh}(x,\omega 
t) d(\omega t) \nonumber\\ 
&=& \frac{J_0}{\omega\pi}(\sin\phi - \phi\cos\phi) 
\end{eqnarray}
From the second of eqns.(21) and the definition of $D_{sh}$, we have:
\begin{equation}
\epsilon_{sh}\frac{d\bar\psi_{sh}}{dx} =  
- \frac{J_0}{\omega\pi}(\sin\phi - \phi\cos\phi) 
\end{equation} 
where we have assumed that the time averaging procedure allows us to 
factor out ~$\epsilon_{sh}$ ~ outside the spatial derivative.
From (19) and (17), we get 
\begin{equation}
\frac{d\phi}{dx} = 
\frac{{\bigl(1-\frac{2}{T_e}\bar\psi_{sh}(x)\bigr)}^{-1/2}}
{s_0\sin\phi} 
\end{equation}
where $\omega t$ ~ was set equal to $\phi$ in (19), ~ $s$ to $x$~, ~~ ~and $s_0=\frac{J_0}{e\omega n_0}$. \\
Combining eqns.(27) and (28) we find 
\begin{equation}
\epsilon_{sh}\frac{d\bar\psi_{sh}}{d\phi} = -\frac{J_0s_0}{2\omega\pi} 
( 1 + \cos 2\phi -\phi\sin 2\phi) 
{\bigl(1-\frac{2}{T_e}\bar\psi_{sh}(x)\bigr)}^{1/2}
\end{equation}
which upon integration leads to
\begin{equation}
{\bigl(1-\frac{2}{T_e}\bar\psi_{sh}(x)\bigr)}^{1/2} = 1 
-\frac{L}{\epsilon_{sh}}\bigl( \frac{3}{8}\sin 2\phi - 
\frac{\phi}{4}\cos 2\phi - \frac{\phi}{2} \bigr) 
\end{equation} 
where 
\begin{equation}
L = \frac{{J_0}^2}{e\pi T_e\omega^2n_0}
\end{equation}
and we have assumed that $\epsilon_{sh}$ is independent of $\phi$. 
Substituting (30) in (17), one finds the following expression for the 
ion density:
\begin{equation}
n_i = n_0 {\bigl\{ 1 - \frac{L}{\epsilon_{sh}}\bigl( \frac{3}{8}\sin 
2\phi - \frac{\phi}{4}\cos 2\phi - \frac{\phi}{2} \bigr) \bigr\}}^{-1} 
\end{equation}
Differentiating (26) with respect to $x$ gives us
\begin{equation}
\nabla\cdot\bar D_{sh} = 
\frac{J_0}{\omega\pi}\phi\sin\phi\frac{d\phi}{dx} = 
\frac{J_0\phi}{\omega\pi s_0}{\bigl\{ 1 - 
\frac{L}{\epsilon_{sh}}\bigl( \frac{3}{8}\sin 2\phi - 
\frac{\phi}{4}\cos 2\phi - \frac{\phi}{2} \bigr) \bigr\}}^{-1}
\end{equation}
Since the net charge density in the sheath $\rho_{sh}$ is given by
\begin{equation}
\nabla\cdot\bar D_{sh} = \rho_{sh}  ~~~,
\end{equation}
one obtains
\begin{equation}
\rho_{sh} = \frac{e\phi n_i}{\pi}
\end{equation}
In order to calculate the sheath capacitance, one must consider the 
instantaneous values. \\
From (23) and (24), the instantaneous displacement field in the sheath 
is 
\begin{eqnarray}
D_{sh} &=& e\int_0^x n_i(\xi)d\xi - \frac{J_0}{\omega}(1-\cos\omega t) 
\nonumber\\ 
&=& \frac{J_0}{\omega}(\cos\omega t - \cos\phi) ~~~, 
~~~~~~~~~~~~~~~~~~~~~ x>s(t)\nonumber\\ 
&=& 0 ~~~~~~~~~~~~~~~~~~~~~~~~~, ~~~~~~~~~~~~~~~~~~~~~ x<s(t) 
\end{eqnarray}
Integrating both sides of (36) with respect to $x$, we get
\begin{eqnarray}
D_{sh}(t) &=& \int_{s(t)}^{s_m} D_{sh}(x,t) dx = \int_{s(t)}^{s_m} 
\epsilon_{sh}E_{sh}(x,t) dx \nonumber\\
&=& \frac{J_0s_0}{\omega}
\int_{\omega t}^{\pi} (\cos\omega t - \cos\phi)\sin\phi \Bigl[ 1 - 
\frac{L}{\epsilon_{sh}}\bigl( \frac{3}{8}\sin 2\phi - 
\frac{\phi}{4}\cos 2\phi - \frac{\phi}{2} \bigr) \Bigr] d\phi \\
&=& \frac{\pi LT_e}{4}(3 + 4\cos\omega t + \cos 2\omega t) - \frac{\pi 
L^2T_e}{\epsilon_{sh}}\Bigl( \frac{15}{16}\pi + \frac{3}{8}\omega t + 
\frac{5}{3}\pi\cos\omega t \nonumber\\
~~&~& + \frac{1}{3}\omega t\cos 2\omega t + \frac{1}{48}\omega t\cos 
4\omega t - \frac{5}{18}\sin 2\omega t - \frac{25}{576}\sin 4\omega t 
\Bigr)
\end{eqnarray}
where a change of variables from ~~$x$ to $\phi$ ~~ has been made in 
(37). We now use these results of the GL theory modified to include a 
finite sheath dielectric constant $\epsilon_{sh}$, to the correct form of Sato's equation (14) for an RF discharge. \\
The contribution to the current in the external circuit from the last 
integral in (14) then is : 
\begin{equation}
\frac{1}{V_a}\int_S \frac{\partial D}{\partial t}\psi_{sh} dS = 
\epsilon_0\frac{1}{V_a}\int_S \frac{\partial E_p}{\partial t}\psi_{sh} 
dS + \frac{1}{V_a}\int_S \frac{\partial D_{sh}}{\partial t}\psi_{sh}dS 
\end{equation}
If it is assumed that the plasma is a good conductor, then the electric field in the plasma can be taken as vanishing. Then it is the last term of (39) whose contribution is relevant to the discharge current.\\ 
We consider again, the GL theory. 
For simplicity, we consider a symmetric discharge, and circular 
electrodes of area $A$ separated a distance $d$ apart. Since the electric displacement and the potential in the sheath are assumed to vary only in the axial direction and are assumed to be uniform in the radial direction (implicit in the assumption that the charge density is uniform across the radius and varies in the axial direction only), therefore, the area can be factored out of the surface integral. If we make also the further assumption that $\epsilon_{sh}(x,t) \approx  \epsilon_{sh}(t)$ ~, that is, that the spatial variation of the sheath dielectric constant can be neglected, then one deduces, making use of (9) and (6) in (39), that the contribution of the term: ~$\frac{1}{V_a}\int_S \frac{\partial D_{sh}}{\partial t}\psi_{sh}dS$  ~ in (40) coming from the RF sheath to the gap capacitance is given by  ~$C_{sh}$ , ~where :
\begin{equation}
C_{sh} = -\frac{A}{V_a}\int D_{sh}(x,t) dx = -\frac{A}{V_a}D_{sh}(t)
\end{equation}
where $D_{sh}$  ~ is given in (38).  Notice, that as expected, this contribution is independent of the distance $d$ separating the electrodes and depends only upon the RF frequency, the electron temperature, the area of the electrodes, $n_0$, the amplitude of the current density, and the sheath dielectric constant $\epsilon_{sh}$.\\   
In the case of RF discharges, the contribution to the total particle flux $\Gamma$ from diffusion of the particles can be taken to be vanishing because the ions and electrons would be expected to react faster to the rapidly changing applied RF voltage, than be influenced by the diffusion gradients in a significant way.
Thus in these cases, only the $n_l{\bf v}_l ~, (l=i, e)$ terms in (3) would contribute to $\Gamma$.\\
The dc contribution to the voltage across the sheath comes from the dc ion current $J_i$. This contributes to the $\Gamma$ ~term in (14), for the current $I$ in the external circuit :
\begin{equation}
J_i = en_0v_B = K{(\frac{2e}{m_i})}^{1/2}
\frac{\bar\psi_{sh}^{3/2}}{s_m^2}
\end{equation} 
where $K = \frac{200}{43}$ in Lieberman's theory, 
while $K = \frac{4}{9}$ for Child's law. $\bar\psi_{sh}$ is given by (30):
\begin{eqnarray}
\bar\psi_{sh} &=& \frac{T_e}{2}{(\frac{L}{\epsilon_{sh}})}^2
{\bigl( \frac{3}{8}\sin 2\phi - \frac{\phi}{4}\cos 2\phi - \frac{\phi}{2}\bigr) }^2 ,  ~~ {\rm for} \phi=\pi \nonumber\\ 
&=&\frac{T_eL^2{(9+2\pi)}^2}{2592\epsilon_{sh}^2}
\end{eqnarray} 
Performing the surface integration:
\begin{equation}
\int_S en_iv_i\psi_{sh} dS 
\end{equation}
over the surface of the discharge space, and assuming as before a simplified geometry of circular electrodes of equal area $A$, and a symmetric discharge, we find that this term contributes
\begin{equation}
\frac{1}{V_a}\int_S e\Gamma\psi_{sh} dS = 
-K{(\frac{2e}{m_i})}^{1/2} \frac{\bar\psi_{sh}^{3/2}A}{s_m^2} \frac{D_{sh}}{\epsilon_{sh}V_a}
\end{equation}
where $D_{sh}$ is given by (38).\\
In going from (29) to (30), we have made the assumption that  $\epsilon_{sh}$ is independent of $\phi$. Since ~$\epsilon_{sh}$ ~ is, in fact frequency dependent:
\begin{equation}
\epsilon_{sh}(\omega) = {\epsilon_{sh}}_\infty + \frac{{\epsilon_{sh}}_s - {\epsilon_{sh}}_\infty}{1 - i\omega\tau}
\end{equation}
where $\tau$ is the sheath relaxation time, it cannot be trivially factored out of the integral. One must use the relation: ~ $\omega t = \phi$ ~ before performing the integration over $\phi$.

\subsection*{Discussion} 
We have shown that the time-dependent version of Sato's equation is not independent of the space charge electric field.  We consider the specific example of the Godyak-Lieberman model for a capacitative RF discharge, and show that the sheath field gives a non-negligible contribution to the gap capacitance of the discharge tube. Space charge effects in a discharge tube can be analysed using energy balance methods, by modelling the system as a two-layer Maxwell-Wagner capacitor. 
For frequencies below RF frequencies also, the argument above shows that the electric field and potential of the space charge can give a non-zero contribution to the external current in the circuit, when the potential drop across the space charge is appreciable. The diffusion terms could also give a non-trivial contribution in these cases to the terms coming from space charge effects :~ $\int_S D\nabla n_l\psi_{sh} dS ~, ~~ (l=i, e) $, ~$D$ being the ambipolar diffusion constant. For a symmetric discharge and circular electrodes of equal area $A$, this surface term would contribute a factor: ~$ \frac{5}{16}\frac{DLn_0}{\epsilon_{sh}^2}{(1 + \frac{3L\pi}{4\epsilon_{sh}})}^{-2} D_{sh}$~~ 
from the ion flux,~~ $D_{sh}$  being given by (38).  
\subsection*{Acknowledgement} 
This work was supported by the Society for Innovation \& Development, 
Indian Institute of Science, Bangalore.\\ 
\newpage
\subsection*{References}
\begin{enumerate}
\item N.Sato, J.Phys.D:Appl.Phys.{\bf 13}, L3(1980).
\item R.Morrow \& N.Sato, J.Phys.D:Appl.Phys.{\bf 32}, L20(1999). 
\item Arthur R. von Hippel, {\em Molecular Science \& Molecular Engineering} (Technology Press of MIT, and John Wiley \& Sons, 1959).
\item M.A.Lieberman, IEEE Trans.Plasma Sci.{\bf 16},638 (1988). 
\item V.Godyak \& N.Sternberg, IEEE Trans.Plasma Sci.{\bf 18}, 159 
(1990). 
\item R.Morrow \& J.J.Lowke, J.Phys.D:Appl.Phys.{\bf 30}, 614 (1997). 
\item B.Chapman, {\em Glow Discharge Processes} (John Wiley \& Sons, 
1980). 
\item S.A.Cohen in {\em Plasma etching : an Introduction}, eds. 
D.M.Manos and D.L.Flamm (Academic Press Inc., 1989).
\item D.Bose, T.R.Govindan, and M.Meyyappan, J.Appl.Phys., {\bf 87}, 
7176 (2000) . 
\item P.A.Miller \& M.E.Riley, J.Appl.Phys.{\bf 82}, 3689 (1997). 
\item W.M.Manheimer, IEEE Trans.Plasma Sci.{\bf 28}, 359 (2000).
\item D.Bohm in {\em The Characteristics of Electrical Discharges in 
Magnetic Fields}, eds. A.Guthrie and R.K.Wakerling (McGraw-Hill, 
1949).
\end{enumerate}
\end{document}